\documentclass[aps,pra,reprint,groupedaddress,floatfix]{revtex4-1}
\usepackage{graphicx}
\usepackage{graphics}
\usepackage{amsmath}
\usepackage{amsthm} 
\usepackage{amssymb}
\usepackage{array}
\usepackage{eqnarray}
\usepackage{mathtools}
\usepackage{dpfloat}
\usepackage{float}
\usepackage{placeins}

\renewcommand{\v}[1]{\ensuremath{\mathbf{#1}}} 
 
\newcommand{\abs}[1]{\left| #1 \right|} 
\newcommand{\pd}[2]{\frac{\partial #1}{\partial #2}} 
 
\newcommand{\ket}[1]{\left| #1 \right>} 
\let\baraccent=\= 
\renewcommand{\=}[1]{\stackrel{#1}{=}} 

\newcommand{\E}{\mathcal{E}}

\begin{document}
\title{Spatial mode storage in a gradient echo memory}

\author{D. B. Higginbottom}
\affiliation{Australian National University}

\author{B. M. Sparkes}
\affiliation{Australian National University}

\author{M. Rancic}
\affiliation{Australian National University}

\author{O. Pinel}
\affiliation{Australian National University}

\author{M. Hosseini}
\affiliation{Australian National University}

\author{P. K. Lam}
\affiliation{Australian National University}

\author{B. C. Buchler}
\email{ben.buchler@anu.edu.au}
\affiliation{Australian National University}

\date{\today}

\begin{abstract}

Three-level atomic gradient echo memory ($\Lambda$-GEM) is a proposed candidate for efficient quantum storage and for linear optical quantum computation with time-bin multiplexing \cite{Hosseini2009}. In this paper we investigate the spatial multimode properties of a $\Lambda$-GEM system.  Using a high-speed triggered CCD, we demonstrate the storage of complex spatial modes and images. We also present an in-principle demonstration of spatial multiplexing by showing selective recall of spatial elements of a stored spin wave. Using our measurements, we consider the effect of diffusion within the atomic vapour and investigate its role in spatial decoherence.  Our measurements allow us to quantify the spatial distortion due to both diffusion and inhomogeneous control field scattering and compare these to theoretical models. 

\end{abstract}


\maketitle

\section{Introduction}

Information processing that harnesses the novel properties of quantum mechanics, such as entanglement and superposition, can be profoundly different and, in some cases, much more powerful than its classical equivalent \cite{Lo2000}. It is this promise that drives the development and implementation of revolutionary quantum communication technologies. Some of the most significant advances in quantum information processing have been made using quantum optics techniques. In particular, optical quantum key distribution (QKD) is already a proven technique for the secure distribution of cryptographic keys via a shared quantum communication channel \cite{Gisin2007,Stucki2002}. However, quantum states are fragile. They are vulnerable to decoherence and measurement processes that destroy information content. This makes the manipulation and storage of quantum information a significant physical challenge. 

To proceed further with optical quantum communication, material systems will be required for the controlled storage and retrieval of quantum light fields. A `quantum repeater' will, for example,  be necessary to extend the range of quantum cryptosystems \cite{Sangouard2011}. Proposed quantum repeater protocols operate by the generation, storage and transfer of entanglement among spatially separated quantum memories. These memories must be capable of coherently storing multiple quantum states of light for on-demand recall with fidelity exceeding the classical limit \cite{Hosseini2011}. In addition, quantum memories are also a key component of proposed linear optical quantum computers \cite{Lvovsky2009}.

The demand for an optical quantum memory has brought forth a host of competing protocols \cite{Lvovsky2009}. Significant progress has been demonstrated in a number of operational systems that couple light fields with atomic ensembles. However, there is as yet no candidate which meets all the benchmarks required for real world applications.

Quantum storage has been demonstrated using electromagnetically induced transparency (EIT) \cite{Fleischhauer2000, Phillips2001}, atomic frequency combs (AFC) \cite{Afzelius2009,DeRiedmatten2008, Chaneliere2010}, and Raman schemes \cite{Reim2011}. Electromagnetically induced transparency (EIT) has achieved efficiencies over $40\%$ \cite{Novikova2008}, been used to store light pulses in a solid state system for multiple seconds \cite{Longdell2005}, and can preserve entanglement  \cite{Zhang2011} and optical squeezing \cite{Akamatsu2004, Appel2007, Honda2008, Arikawa2009}. Entanglement has also been stored in AFC memories \cite{Saglamyurek2011, Clausen2011}, which have the convenience of large signal bandwidths (several GHz) \cite{Bonarota2011}. 

In this work our quantum memory is a warm-gas gradient echo memory (GEM). GEM, which is a variant of the controlled reversible inhomogeneous broadening (CRIB) protocol \cite{Moiseev2001}, is a photon echo memory technique that uses applied electric or magnetic fields to reverse the time evolution of an atomic coherence in an inhomogeneously broadened sample. Coherent light storage has been achieved in room temperature rubidium vapour with efficiencies as high as 87\% using this method \cite{Hosseini2011}. State-independent verification using the conditional variance and signal transfer coefficient has shown unambiguously that the memory performs beyond the quantum no-cloning limit  \cite{Hosseini2011a}. 

\begin{figure*}[t]
\begin{center}
\includegraphics[width=5.8in]{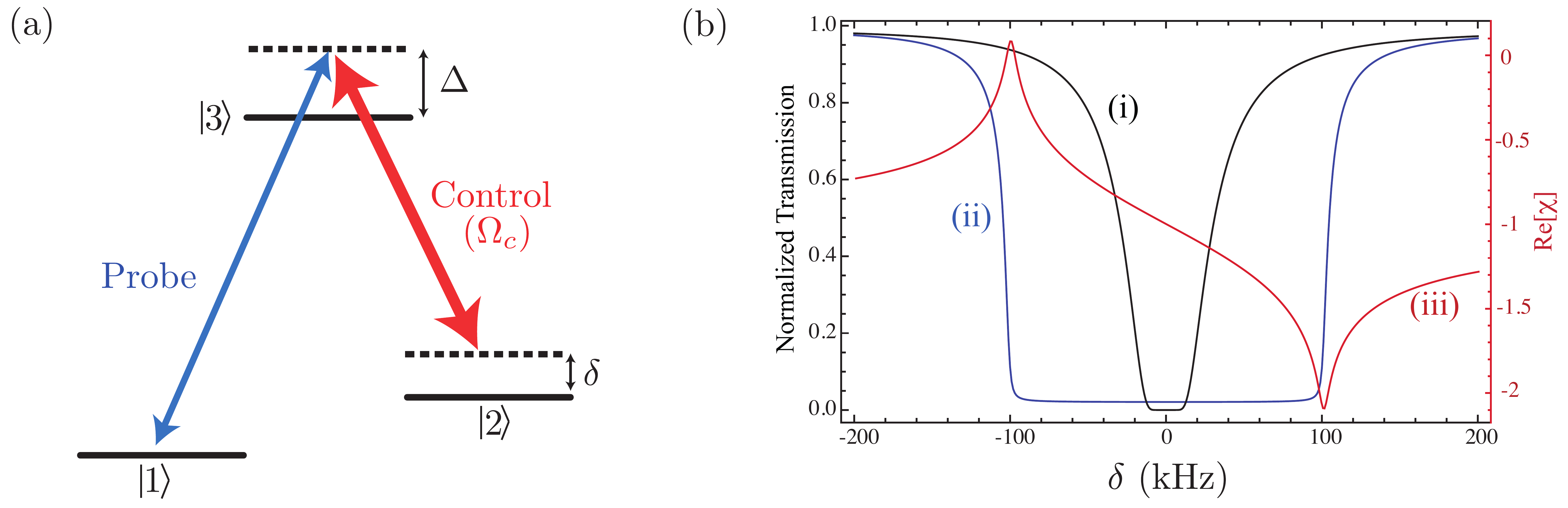} 

\caption{(a) The off resonance Raman level scheme used for $\Lambda$-GEM. The probe beam is detuned from the transition $\ket{1}\rightarrow \ket{3}$ by $\Delta$, the single photon detuning. The control beam is detuned from the transition $\ket{3} \rightarrow \ket{2}$ by $\Delta_c$. The detuning relevant to the Raman transition is the two photon detuning $\delta = \Delta - \Delta_c$. (b) The properties of the ensemble as a function of the two photon detuning $\delta$. (b-i) The unbroadened Raman absorption feature, the medium is most opaque to light that is resonant with the Raman transition $\delta = 0$. (b-ii) The inhomogeneously broadened Raman feature of the ensemble. The magnetic field makes the detuning a function of longitudinal position in the ensemble, the absorption feature is broadened because a wider range of frequencies are resonant with some component of the ensemble. (b-iii) The real component of the ensemble electric susceptibility $\chi$. The dispersion is greatest at the edges of the broadened Raman feature.}\label{levelscheme}
\end{center}
\end{figure*}

In this paper we examine the spatial multimode properties of a GEM. Multimode capacity describes the number of optical modes (spatial, temporal or frequency) that can be stored in a memory. A quantum memory with the capacity to simultaneously store orthogonal spatial modes is valuable because it allows parallel storage and processing of multiple signals- a single-cell device for multi-qubit memory. Repeaters composed of multimode memories can increase the channel bit-rate dramatically by multiplexing between modes \cite{Collins2007,Simon2007,Vasilyev2008}. In addition, quantum correlated images may themselves form the basis of new quantum information protocols \cite{Pooser2008, Marino2009}. A quantum memory with high spatial fidelity is necessary for the storage of such signals. 

Previous work has shown storage of images using EIT \cite{Vudyasetu2008,Shuker2008,Heinze2010,Ding2012} and four-wave mixing \cite{Marino2009,Wu2012}. We also note very recent work demonstrating the storage and recall of consecutive images in a $^{85}$Rb vapour using GEM \cite{Glorieux2012}. In this work we investigate the capacity of a gradient echo memory to store complex spatial modes and multimode images, and measure the deterioration of spatial fidelity and recall efficiency as a function of storage time. We also perform an in principle demonstration of in-memory spatial processing. Finally, the spatial fidelity of our quantum memory is an important diagnostic tool, by examining the deterioration of the storage efficiency for complex spatial modes we probe the impact of diffusion in the vapour cell memory.

\section{Three level GEM}

The quantum memory used in these experiments is a three level ($\Lambda$) GEM which stores optical information in the long-lived coherence between hyperfine ground states of warm $^{87}$Rb atoms. The key advantage of three-level memories is that they harness the long coherence times between negligibly coupled ground states \cite{Balabas2010}.

A strong classical control beam couples the probe signal to the two ground states $\ket{1}$ and $\ket{2}$ via an excited state $\ket{3}$ in an off-resonance Raman configuration illustrated in Fig.~\ref{levelscheme}a. For a weak probe beam and large single photon detuning ($\Delta$) the excited state population is negligible during the storage process. In this way the probe signal is coherently transferred to a spin coherence  $\rho_{12}(\v{r},t)$ in the ensemble. The control field needs to be on for read and write operations, but can remain off during storage.

The Zeeman sub-level splitting of the atomic medium can be controlled using magnetic fields. We use two solenoids to produce `read' and `write' fields with linear gradients $\eta$ and $-\eta$ along the signal propagation axis of the atomic medium. These fields inhomogeneously broaden the resonant frequency of the Raman transition producing the absorption feature shown in Fig.1b. Individual frequency components of the input are mapped  longitudinally along the cell to produce a spatial spin wave corresponding to the Fourier spectrum of the input field envelope. 

Once the wave-packet is stored, the atomic dipoles precess with an angular velocity proportional to their local resonant frequency. Over time the spin excitation accrues a spatial phase variation which makes coherent re-emission impossible. The signal can be retrieved by reversing the system's time evolution. Switching the magnetic field gradient from $\eta$ to $-\eta$ at some time $\tau$ after storage inverts the local detuning and causes the dipoles to rotate in the opposite direction. At time $2\tau$ the dipoles realign and the atomic spin wave is once again in phase. A time reversed echo of the original signal will be released in the forward direction. GEM can operate with recall efficiency approaching unity in the forward direction \cite{Hetet2008c, Longdell2008}.

\begin{figure*}[t]
\begin{center}
\includegraphics[width=6.6in]{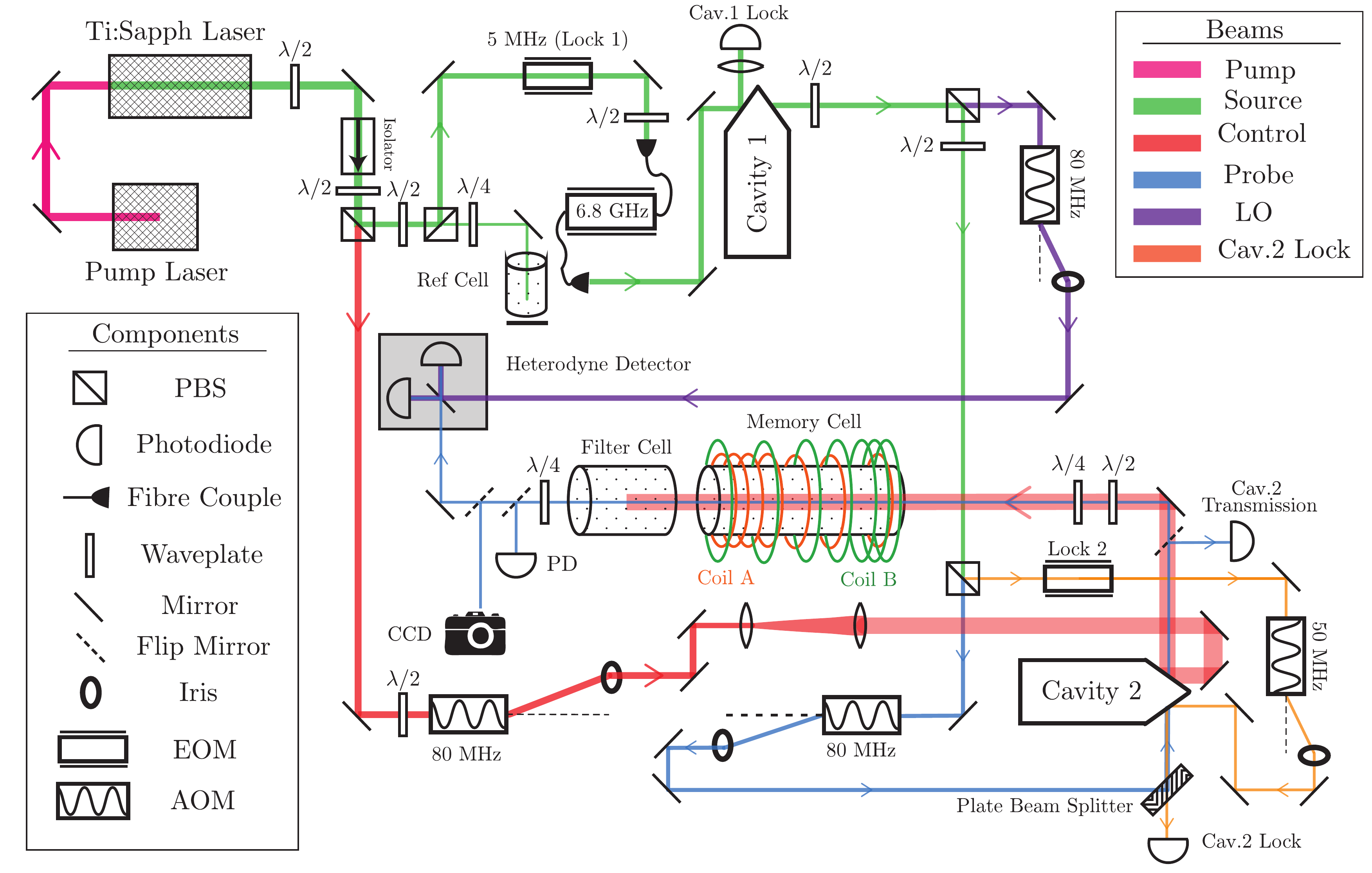}
\caption{Experiment schematic, described in text. PBS: polarising beam splitter, EOM: electro-optic modulator, AOM: acousto-optic modulator, CCD: charge coupled device (camera).} \label{expschematic}
\end{center}
\end{figure*}

In the $\Lambda$ configuration pulse emission can only take place if all atomic dipoles oscillate in phase while the Raman control beam is switched on. Leaving the control field off suppresses the echo and the signal remains stored in the atomic excitation. This condition can be exploited to control the order in which stored pulses are retrieved, thus enabling resequencing of time-bin qubits \cite{Hosseini2009}. Furthermore, the applied magnetic fields make in-memory spectral manipulation possible \cite{Buchler2010}. These degrees of control readily extend to spatial processes and make $\Lambda$-GEM a promising candidate for spatially multiplexed memories.

\section{Experiment}

The essential schematic of this experiment is illustrated in Fig.~\ref{expschematic}, it shows the configuration of the beams, modulators, cavities and detectors that prepare our gradient echo memory. The probe and control beams are derived from a single continuous wave Ti:Sapph laser source which produces 1.1 W of power at 795 nm. The laser output is tuned to the $^{87}$Rb D1 transition ($F = 2 \rightarrow F$'$ = 2$) using an external rubidium reference cell.  The source laser is then blue detuned from this transition by approximately 1.5 GHz by monitoring the reference cell fluorescence.

To produce the probe we transfer the source beam power into FM sidebands at $\pm 6.8$ GHz with a fibre coupled EOM. We use a ring cavity (Cavity 1 in Fig.~\ref{expschematic}) to select the $+6.8$ GHz sideband from the modulated signal. This sideband frequency corresponds to the ground state splitting of $^{87}$Rb.  

We operate a second ring cavity (Cavity 2 in Fig.~\ref{expschematic}) as a spatial mode cleaner, it transmits the probe light in only one spatial mode selected by cavity alignment. The output port of the cavity is also used to combine the probe and control beams.

The memory cell is an anti-reflection (AR) coated pyrex cylinder $200$ mm long and $25$ mm diameter containing a mixture of isotopically enhanced $^{87}$Rb and $0.5$ Torr of krypton buffer gas. The Raman transition between hyperfine states is inhomogeneously broadened by magnetic coils that apply a linearly varying Zeeman shift along the propagation axis of the vapour cell. Current is switched between two solenoids of opposite pitch to create magnetic fields for the read and write operations (coils A and B in Fig.~\ref{expschematic}). A third coil (not shown) provides an adjustable constant field offset.

After the memory, the control beam is filtered from the signal by absorption in a second gas cell containing a natural mixture of rubidium isotopes. The vapour pressure in the memory and filter cells is fixed by adjusting the temperature with electrical elements. The memory cell is kept at $70^\circ$C  and the filter cell is kept at approximately $140^\circ$C.

Once the filter cell has suppressed the control beam at the memory output, flip mirrors can be used to switch the signal between three detection mechanisms. In the first a local oscillator beam in a heterodyne configuration is used to perform phase or amplitude measurements on the signal retrieved from the memory. In the second a photodiode can be used to record the temporal intensity of stored pulses independent of spatial mode. Lastly, we can send the beam to a high-speed CCD camera for spatial mode analysis.

The Grasshopper2 CCD camera from Point Grey Research features a Sony ICX285 $1.4$ megapixel image sensor with resolution 1384 x 1036. The total size of the CCD sensor is $20 \times 14$ mm. Each image is the average of fifty $30$ $\mu$s exposures triggered externally from a digital control station. Alternating images of the echo and background are taken so that background signal (including control beam leakage) can be removed dynamically. The subtracted background is the mean of two no-signal images taken before and after the echo. The response of the CCD was calibrated against the response of the photodiode to ensure that the integrated CCD measurements gave an accurate measure of total energy in the pulse.

The cavity locking, probe and control beam intensities, magnetic field configuration, heater and camera triggers are controlled by a script specially written in LabVIEW$\textregistered$ \cite{Sparkes2011a}.

\section{Results}

\subsection{Fundamental mode efficiency}

Storage time is a critical parameter for quantum memory systems and it is limited by multiple decoherence mechanisms. Ground state decoherence, collisional broadening, scattering processes and diffusion all contribute to the relaxation of the atomic spin wave. To improve the achievable storage time it is vital to understand the mechanisms behind the loss of efficiency and fidelity.

A spatial mode investigation gives us a new window to the operation of our memory. In this first section we compare heterodyne and CCD measurements of the recall efficiency of TEM-00 echoes. The CCD, unlike the heterodyne detector, has neither fine temporal resolution nor spatial mode sensitivity. It detects all spatial modes and frequencies and integrates the signal over an exposure time of $30$ $\mu$s. However it allows us to collect spatial information and to operate with complex spatial profiles which are not accessible to the heterodyne system. In the following sections we will use the CCD to investigate the storage of both high-order Hermite-Gauss spatial modes and multimode images.

Under certain conditions degenerate four-wave mixing (FWM) between probe and control fields configured as per Fig.~\ref{levelscheme}a and a third conjugate field can be expected to cause amplification of the probe signal  \cite{Hosseini2012b}. The phase matching condition of the FWM process imposes restrictions on the propagation of the generated conjugate field so that when linearly polarised coupling and probe fields propagate through the medium at a non-zero crossing angle we observe the conjugate field as a distinct spatial mode at the CCD camera. For this experiment we use circularly polarised light and operate at a temperature and one-photon detuning such the conjugate field intensity and the associated probe amplification are negligible. By this method we ensure that the results presented here are free from efficiency or spatial mode distortion due to FWM.

\begin{figure}[]
\begin{center}
\includegraphics[width=3.3in]{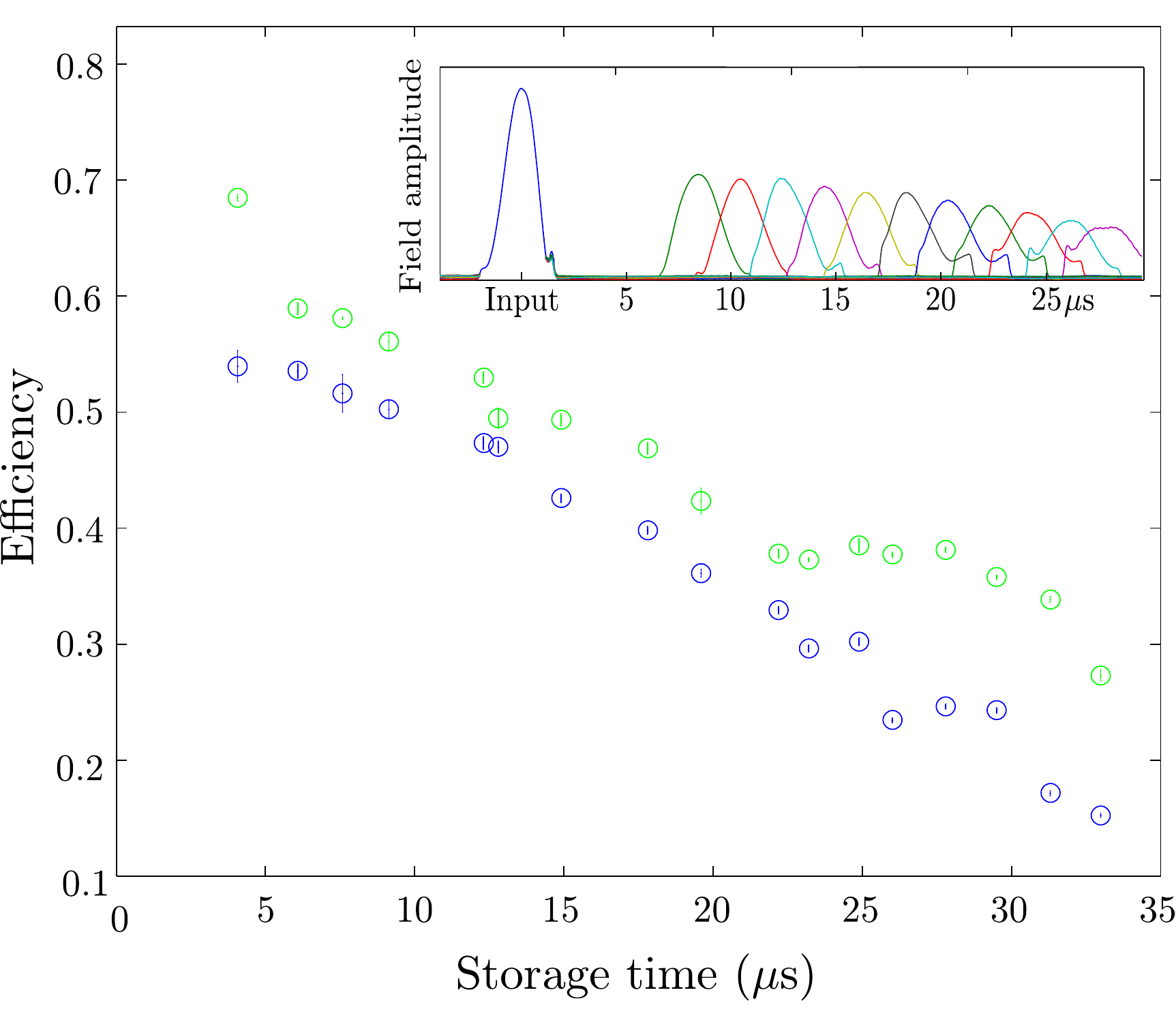}
\caption{Comparison of TEM-00 storage efficiency measured by heterodyne detection (blue) and CCD camera (green). The heterodyne efficiency is the area under the echo pulse compared to the area of the input. The CCD efficiency is the total intensity on the camera compared to the input. Inset: Demodulated heterodyne traces showing the shape of the photon echo at several recall times.} \label{comparehetcam}
\end{center}
\end{figure}

The efficiency with the control field both on and off during storage has previously been characterised using heterodyne detection \cite{Hosseini2011}, but this method will report extra losses if components of the echo have changed mode. The inset in Fig.~\ref{comparehetcam}  shows the temporal profile of TEM-00 echoes as a function of storage time. For higher order spatial modes we used the photodiode detector (which is not mode sensitive) to check the temporal shape of the recalled pulse. The time profiles of the higher order mode echoes were not more distorted than the fundamental mode traces shown in Fig.~\ref{comparehetcam}.  

The results of efficiency measurements made using the heterodyne and CCD are compared in Fig.~\ref{comparehetcam}.  The control beam is left off during storage to minimise control scattering losses (see Section IV-C). The efficiency measured by the CCD tracks above the concurrent heterodyne efficiency and the difference increases with storage time. The difference is light which reaches the detector in a spatial mode orthogonal to the local oscillator or has spread outside of the local oscillator beam. The increasing gap suggests that there are processes in the memory that disperse spatial information during storage.

\subsection{Diffusion of TEM-00}
 
  \begin{figure*}[t]
\begin{center}
\includegraphics[width = 6.6in]{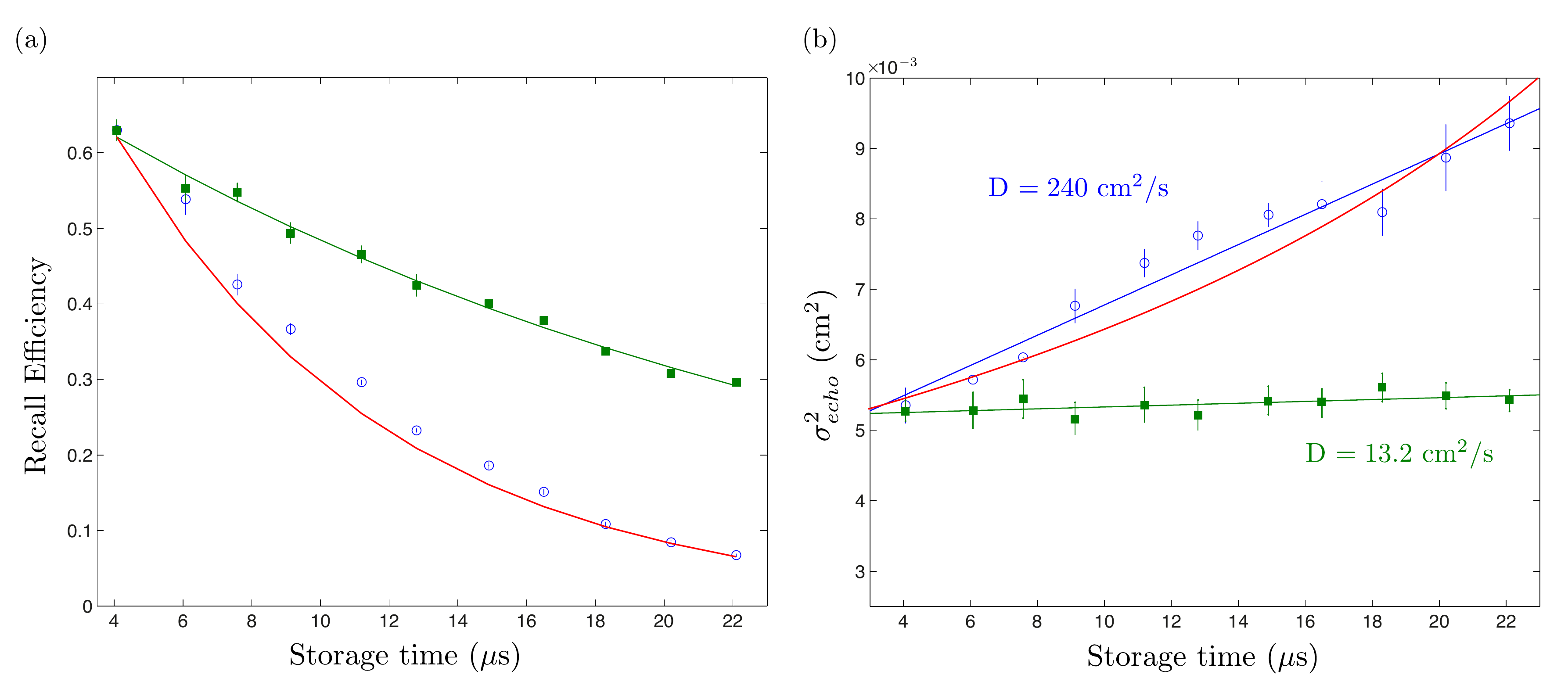}
\caption{(a) Recall efficiency of control-off (green squares) and control-on (blue circles) echoes as a function of storage time from the CCD. The solid red line is a prediction of the control-on echo efficiency from the additional decoherence due to control field scattering (Section IV-C). (b) Area of the Gaussian mode echo as a function of storage time from the same data. The points plotted are $\sigma^2$ where $\sigma$ is the standard deviation of the echo intensity distribution. Data is shown for both control-on (blue circles) and control-off (green squares) storage along with solid lines showing a linear fit to each dataset. The gradient of this fit corresponds to the diffusion coefficient $D$ implied by the mode expansion (shown on plot). The red line is a theoretical prediction for the control-on width from simultaneous diffusion and inhomogeneous control field scattering (Section IV-C).}\label{tem00area}
\end{center}
\end{figure*}

The dominant spatial effect in warm vapour memories is atomic transport by Brownian motion. It has been shown that the ballistic motion of warm atoms can coherently distribute a collective excitation throughout a gas cell \cite{Xiao2008}. 
Therefore we expect to measure expansion of warm GEM photon echoes corresponding to diffusion of the atomic excitation in the memory.

Under certain conditions the internal atomic degrees of freedom are essentially decoupled from the atom's external motion \cite{Xiao2008}, and the two may be considered separately. The exceptions to this principle are atom-atom and atom-wall collisions, field inhomogeneity, Doppler broadening and other velocity effects \cite{Firstenberg2008}. In $\Lambda$-GEM memories the field is homogeneous on the plane transverse to the solenoid axis. Displacement on this plane does not change the internal state of each atom, therefore the atomic motion can effectively be modelled by the addition of a classical diffusion term $\dot{\rho}_\mathrm{diff} = D \nabla^2 \rho$ to the Maxwell-Bloch equations for the collective atomic states where $D$ is the diffusion coefficient of the vapour \cite{Firstenberg2008}. The expectation values of the probe field envelope $\E$ and spin coherences $\rho_{12}$ and $\rho_{13}$ evolve according to \cite{Hetet2008c}

\begin{align}\label{maxbloch}
\pd{\E}{z} & =    \frac{i g n}{c} \rho_{13} + \frac{i}{2 k_0} \nabla_{x,y}^2 \E \\
\dot{\rho}_{13} & = ig\E + i\Omega_c \rho_{12} - \frac{1}{2} \left( 2\gamma + \gamma_0 + \gamma_c \right) \rho_{13} + i\Delta \rho_{13} \nonumber \\
&  + D \nabla^2 \rho_{13} \nonumber \\
\dot{\rho}_{12} & = i\Omega_c \rho_{13}  - \left( \gamma_0 + \gamma_c - i\delta +\frac{i\Omega_c^2}{\Delta} \right) \rho_{12} + D \nabla^2 \rho_{12} \nonumber
\end{align}

Where $g$ is the vacuum Rabi frequency of the probe light mode, $n$ is the rubidium density in the cell, $k_0$ is the wavenumber of the probe mode and $\gamma$, $\gamma_0$ and $\gamma_c$ are the excited state decay, dephasing and population exchange rates respectively. The first equation has been simplified by transforming into a frame moving at the speed of light along the longitudinal axis $z$.

In GEM longitudinal and transverse diffusion must be considered separately because of the applied longitudinal field gradient. We can model transverse diffusion with a Gaussian propagator. When the atomic mean free path is much smaller than the radius of the bounding region, the solution for the ensemble atomic spin wave $\rho(\v{r},t)$ far from the cell boundaries is the convolution of the solution in the absence of diffusion $\rho_\mathrm{stat}(\v{r},t)$ with a diffusion propagator $G(\v{r},t)$ of width $\sigma_\mathrm{diff} = \sqrt{2 D t}$ \cite{Shuker2008}
 
 \begin{equation}\label{diffusionsolution} 
 \begin{array}{ccl}
 G(\v{r},t) &=& \left(4 \pi D t\right)^{-N/2}  \exp{\left(-\frac{\v{r} \cdot \v{r} }{2\sigma_{\mathrm{diff}}^2}\right)}\\
\rho(\v{r},t) &=& \int d\v{r}'  G(\v{r} - \v{r}' , t) \rho_\mathrm{stat}(\v{r}', 0)
\end{array}
\end{equation}

Where $N = 2$ is the number of dimensions relevant for transverse diffusion. The recalled field is determined by the locally averaged value of all displaced atoms within a small region. That is, by the diffused atomic coherence operator above. The recalled field envelope mapped from the diffused coherence is

\begin{equation}\label{recallsolution}
\abs{E(\v{r},t)}^2 = \abs{P(t) \int d\v{r}' G(\v{r} - \v{r}' , t) E(\v{r},0)}^2
\end{equation}

Where we have introduced a homogeneous power loss function P(t) which includes the profile's exponential decay due to scattering and decoherence as well as losses caused by longitudinal diffusion which are not in general exponential. For a Gaussian beam (TEM-00) with a uniform phase front and waist $W_0$ this diffusion model causes the total observed intensity to drop like

\begin{equation}\label{powerdrop}
\int \abs{E(\v{r},t)}^2 d\v{r} = \frac{W_0^2}{4Dt + W_0^2} \int \abs{E(\v{r},0)}^2 d\v{r}
\end{equation} 

Figure~\ref{tem00area}a shows the decay of the recalled energy as a function of storage time with and without the control field left on during storage. Previous work has shown that control-off storage is more efficient as it reduces the amount of spontaneous Raman scattering \cite{Hosseini2011}. Our model accounts for a spatially inhomogeneous control field and  predicts precisely the additional losses from the control field.

Using the CCD we are now able to measure the effect of the control beam on the mode profile of the recalled echo. Defining $\sigma$ as the standard deviation of average beam profile, Fig.~\ref{tem00area}b shows the area ($\sigma^2$) of a 2D Gaussian fit to the recalled pulse image both with and without the control beam left on during storage.

Convolving the TEM-00 profile with a Gaussian diffusion propagator produces a broadened Gaussian with width $\sigma_{\mathrm{echo}} = \sqrt{\sigma_{\mathrm{in}}^2 + \sigma_{\mathrm{diff}}^2}$). Therefore the measured echo intensity profile has width $\sigma_\mathrm{echo}^2/2 =  W(z)^2/4 + \sigma_\mathrm{diff}^2/2$ where W(z) is the beam width of the input mode. This allows us to infer the diffusion coefficient from the measured echo expansion.

\begin{equation}\label{inferredD}
D = \pd{ ( \sigma^2)}{t} 
\end{equation}

The diffusion coefficient is related to the collision rate $\gamma_\mathrm{coll}$, mean free path $\lambda$ and atomic velocity $v$ by 

\begin{equation}\label{dcoefficient}
D = \lambda \bar{v}/3 = \bar{v}^2 / 3\gamma_\mathrm{coll}
 \end{equation}

The collision rate is well known for a number of low pressure binary mixtures, therefore we can estimate theoretically the diffusion coefficient within the rubidium cell. For low pressure rubidium in krypton buffer gas the collision rate is approximately 17 MHz/Torr of buffer gas pressure at room temperature \cite{Happer1972}. This implies that $D \approx 31$ cm$^2$/s in our memory vapour cell.

The expansion rate calculated with the control beam on (Fig.~\ref{tem00area}b) is $240 \pm 18$ cm$^2$/s, a factor of eight higher than the expected atomic diffusion rate in the cell. It is clear that the control beam is causing additional distortion of the profile during storage. We explore this mechanism in the following subsection, and show that control beam scattering provides a good explanation for the control-on TEM-00 expansion data.

In contrast when the control beam is off we measure only half the expansion expected from diffusion. In the control-off data the rate of mode expansion corresponds to a diffusion coefficient $D = 13.2 \pm 1.6$ cm$^2$/s. To resolve this discrepancy we performed a direct, independent measurement of diffusion in the cell using the method of Gozzini and Bicchi \cite{Gozzini1967,Bicchi1980} which indicated a diffusion coefficient of $D = 65 \pm 10$ cm$^2$/s. This appears to confirm that the measured expansion of the recalled echo signal is smaller than the expansion expected from atomic diffusion.

Echo mode expansion is only an indirect measurement of diffusion in the cell.  Indeed, there are mechanisms that could cause a discrepancy between the diffusion rate and the measured mode expansion rate. For example, magnetic field inhomogeneity in the transverse dimension could reduce rephasing efficiency at the edges of the diffusing spin wave. Numerical simulations show  that diffusion in the presence of transverse magnetic field variation does cause additional echo efficiency loss and reduced echo mode expansion. However, measurements of the field variation in our setup indicate field fluctuations that are orders of magnitude too small to explain the observed discrepancy, so we can rule this mechanism out as a major factor. Another possible explanation of the smaller than expected beam size is a reduction of the beam divergence due to diffusion \cite{Firstenberg2010}. The beam waist, which lies in the cell, can expand due to diffusion leading to a beam with reduced divergence that when measured downstream will appear smaller than otherwise expected.

\subsection{Control field scattering}

 \begin{figure}[t]
\begin{center}
\includegraphics[width=3.3in]{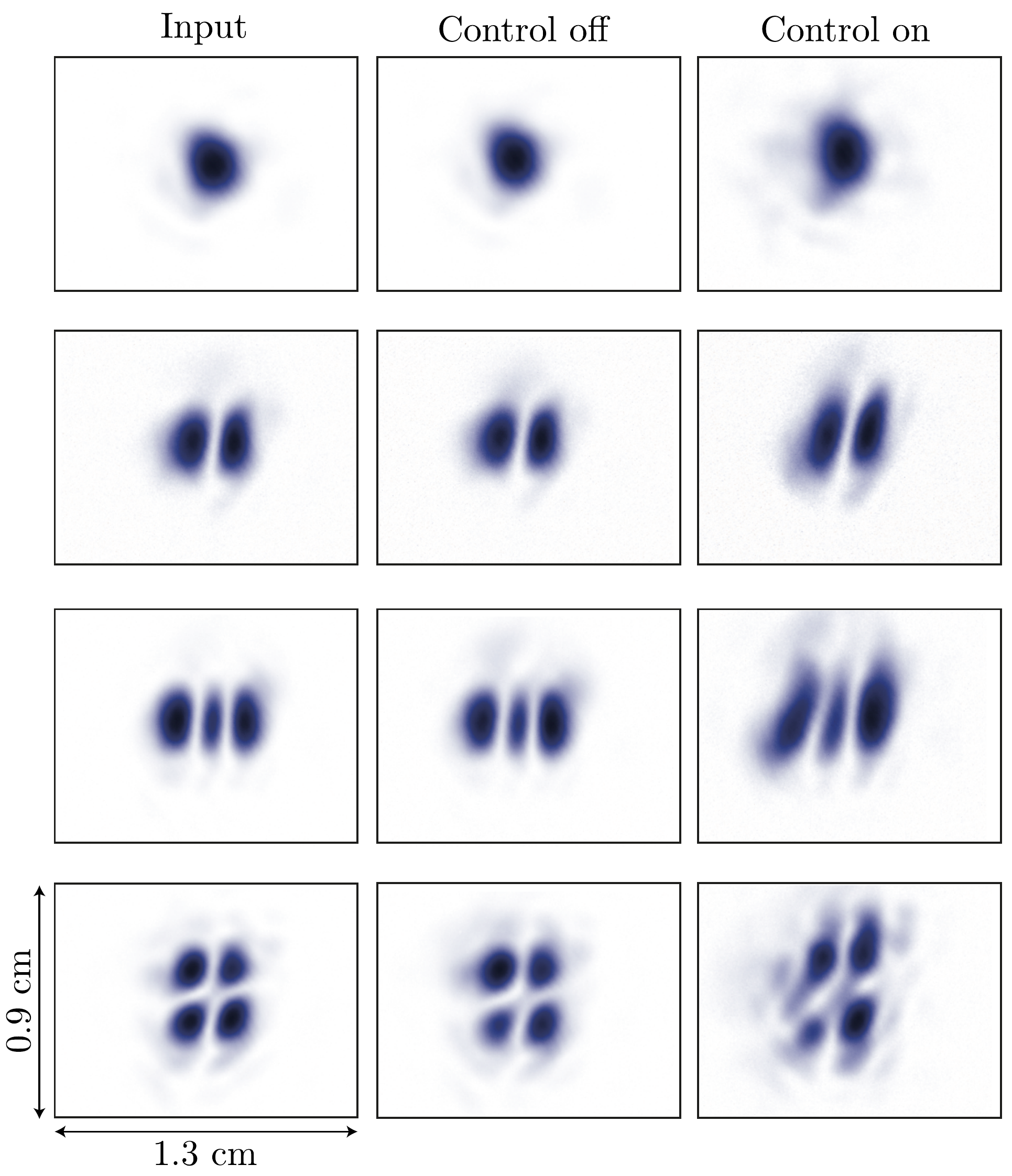} 
\end{center}
\caption{The spatial profile of several Hermite-Gauss mode inputs (left) and photon echoes stored for $12$ $\mu$s with the control beam off (middle) and on (right) during storage. The image intensities have been normalised, i.e. they do not show the overall decrease in recall efficiency.}\label{controlonoffpics}
\end{figure}

\begin{figure*}[t]
\begin{center}
\includegraphics[width=6.3in]{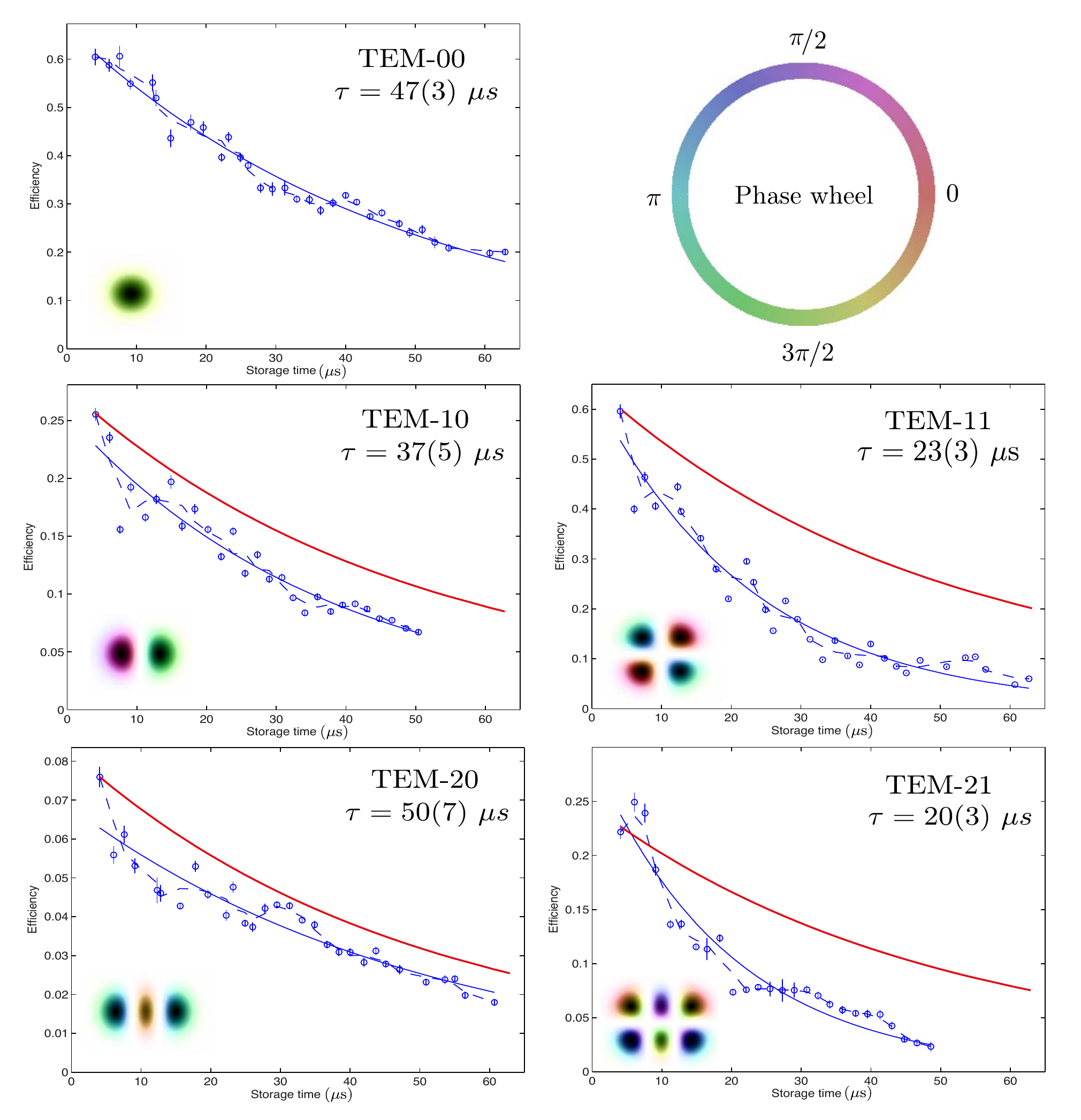} 
\end{center}
\caption{Recall efficiency of the Hermite-Gauss TEM modes with control off during storage. Control window clipping adds experimental noise in short cycles. The dotted line shows a three point mean filter to remove this noise. The blue trace is an exponential fit to the data of the form $Ae^{-t/\tau}$. The decay time $\tau$ from this fitting is shown on each plot. The red curve is the TEM-00 decay curve shown for comparison. }\label{TEMnm}
\end{figure*}

With our imaging experiment we can see that in addition to the overall loss of efficiency the spatial profile of the recalled pulses is significantly worse when the control has been left on during storage. Figure~\ref{controlonoffpics} compares the recalled profile of several Hermite-Gauss modes with and without the control beam during storage.

A significant proportion of this distortion can be explained by the inhomogeneity of the control beam. As the control field power varies across the Gaussian profile so does the control field two-photon scattering rate $\Gamma$. 

\begin{align} \label{scatteringrate}
\Gamma & =  \gamma_{31} \frac{\Omega_c^2}{\gamma_{31}^2 + \Delta^2} \exp \left( \frac{-2 (\v{r} \cdot \v{r})}{W_c^2}\right)  \\ & \approx \gamma_{31} \left(\frac{\Omega_c}{\Delta}\right)^2 \exp \left( \frac{-2 (\v{r} \cdot \v{r})}{W_c^2}\right) \hspace{5mm}  \mathrm{for} \hspace{5mm} \Delta >> \gamma_{31} \nonumber
\end{align}

Where $\gamma_{31} = 2\pi \times 5.6$ MHz is the excited state decay rate \cite{Steck2001}, $\Delta = 1.5$ GHz is the one photon detuning and $\Omega_c$ is the control field Rabi frequency. For our control beam power (400 mW) and waist ($W_c = 3$ mm) the Rabi frequency is $\Omega_c = 72$ MHz. Spatially dependent scattering will burn out features from the probe signal in regions of higher control intensity. The control beam has a Gaussian intensity distribution which will flatten the recalled pulse profile, leading to an increase in the observed width.

In Fig.~~\ref{tem00area}a we have plotted the TEM-00 recall efficiency (measured by the CCD) as a function of time for control-on and control-off storage. The red curve is the control on efficiency expected from additional exponential decay at the control field scattering rate $\Gamma$. Control field scattering accounts very precisely for the additional loss of efficiency, and is the dominant loss mechanism when the control beam is on. 

Furthermore, we modelled the impact of simultaneous diffusion and control field feature burning and found that this accounts for the broadening in the TEM-00 echo profile when the control beam is on. The red curve in Fig.~\ref{tem00area}b is the predicted expansion from simultaneous diffusion and inhomogeneous control field scattering at the rate $\Gamma$. The model featured perfectly aligned co-propagating control and probe beams with respective widths $W_p = 1.5$ mm and $W_c = 3$ mm. The diffusion constant used in this model was the rate implied by the expansion of the control off TEM-00 mode, $ D \approx 13$ cm$^2$/s, but the broadening due to diffusion in this case is negligible compared to the broadening from the control field.

\subsection{Hermite-Gauss mode efficiency}

In the absence of atomic motion and control beam inhomogeneity we expect all spatial profiles to behave identically in the memory. Simulations performed with three-dimensional Maxwell-Bloch equations (equation~\ref{maxbloch}) confirm that under these conditions $\Lambda$-GEM is insensitive to spatial mode. However, in the presence of atomic motion there are reasons to expect complex spatial modes to decohere more rapidly during storage than a simple Gaussian profile \cite{Shuker2008, Zhao2008a, Wang2008}.

During the memory write operation the transverse spatial profile of the electric field is transferred to the atomic excitation. This preserves spatial phase by creating a spin wave that mirrors the optical field envelope. When atomic transport is possible, atoms may drift between out of phase regions of the excitation. Mixing components of the spin wave in this way reduces the average coherences $\rho_{12}$ and $\rho_{13}$ and prevents efficient rephasing of the photon echo.

\begin{figure}[t]
\begin{center}
 \includegraphics[width=3.4in]{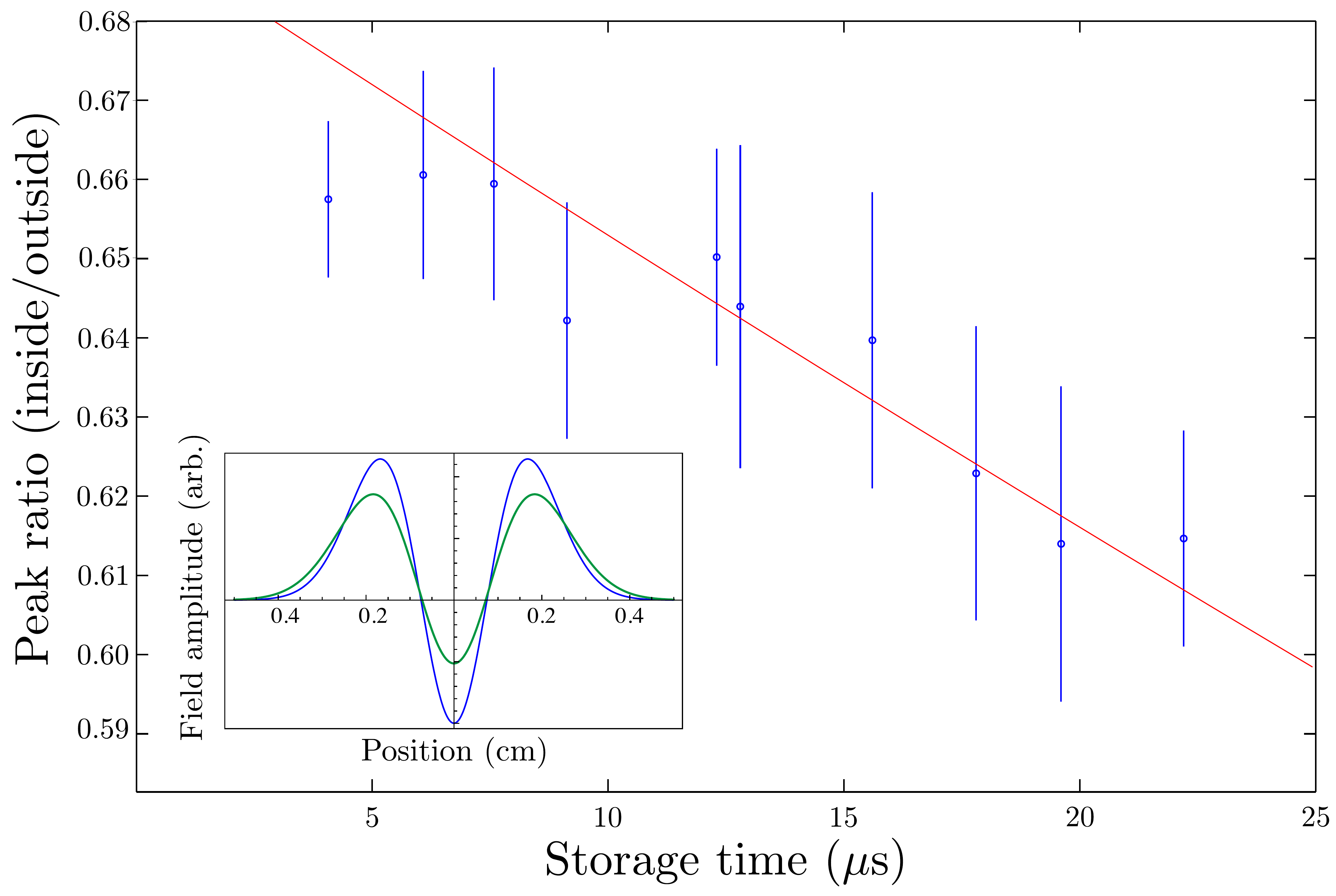}
\end{center}
\caption{The inside/outside peak height ratio of control-off TEM-20 photon echoes as a function of storage time. Diffusion between the out of phase peaks causes the intensity to decrease fastest at the centre of the profile. The large uncertainties are due to fluctuations in the relative heights between images. The red line is a diffusion model with $D \approx 13$ cm$^2$/s and also includes inhomogeneous control field scattering during the read and write operations. Inset: spatial profile of TEM-20 before (blue) and after (green) distortion induced by our diffusion model.}\label{peak height}
\end{figure}

The set of Hermite-Gauss spatial modes produced by our second ring cavity is ready-made for the investigation of spatial spin wave interference. We stored multiple Hermite-Gauss modes  for up to 60 $\mu$s and measured the recall efficiency of each mode as a function of storage time. We found that the decay rate is correlated to the phase complexity of the spatial profile. Figure~\ref{TEMnm} shows the efficiency loss curves of four higher order Hermite-Gauss modes TEM-mn compared to the efficiency of the Gaussian mode TEM-00. The echoes were stored with the control beam off and the control beam alignment was optimised separately for each mode. The magnetic fields were kept consistent between experiments. Each curve is shown with a profile of the corresponding spatial mode in which darkness corresponds to field intensity and hue indicates the field phase as given by the phase wheel at top right.

\begin{figure*}[t]
\begin{minipage}[b]{0.48\linewidth}
\centering
\includegraphics[width=2.1in]{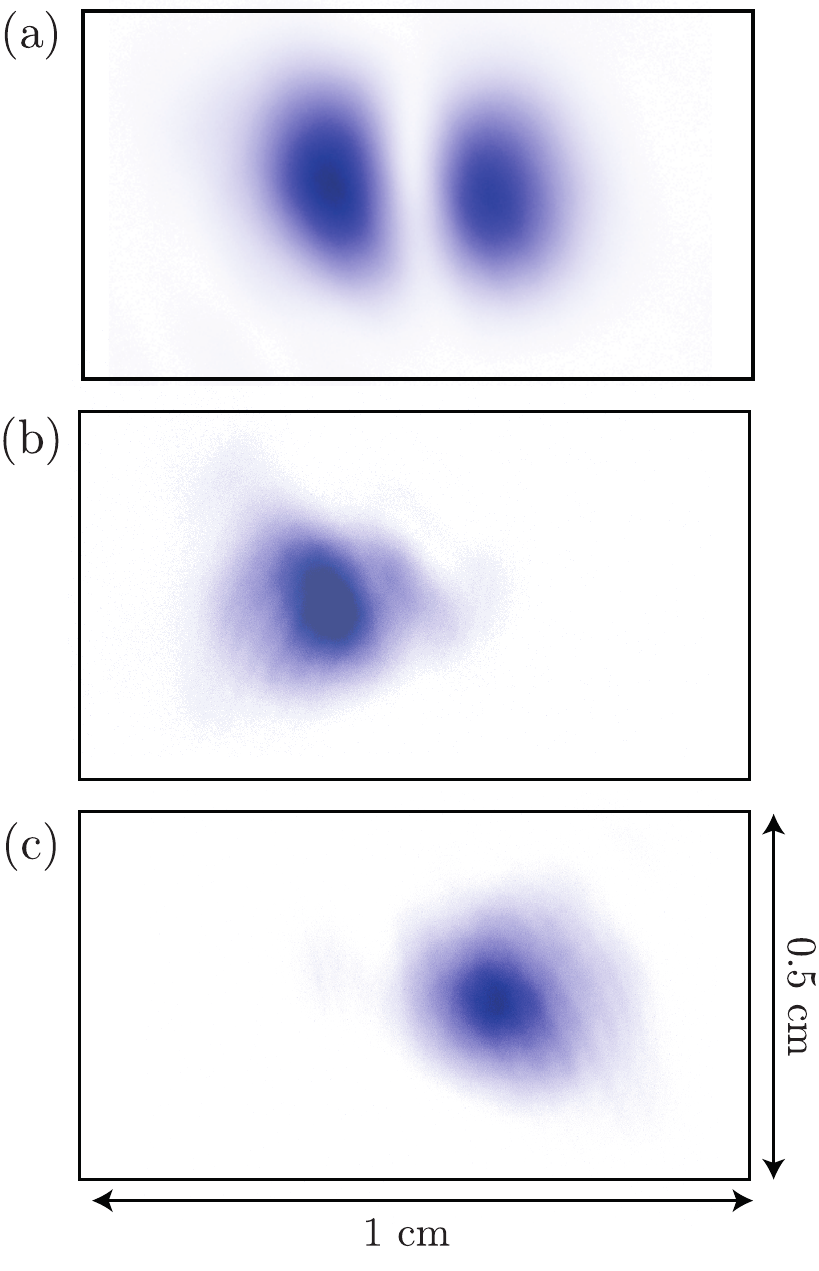}
\caption{(a) Input profile of the TEM-10 mode  (b) The left half of the profile, recalled after $6$ $\mu$s storage. (c) The right half of the profile, recalled after the same time.}\label{wink}
\end{minipage}
\hspace{0.5cm}
\begin{minipage}[b]{0.48\linewidth}
\centering
\includegraphics[width=3.3in]{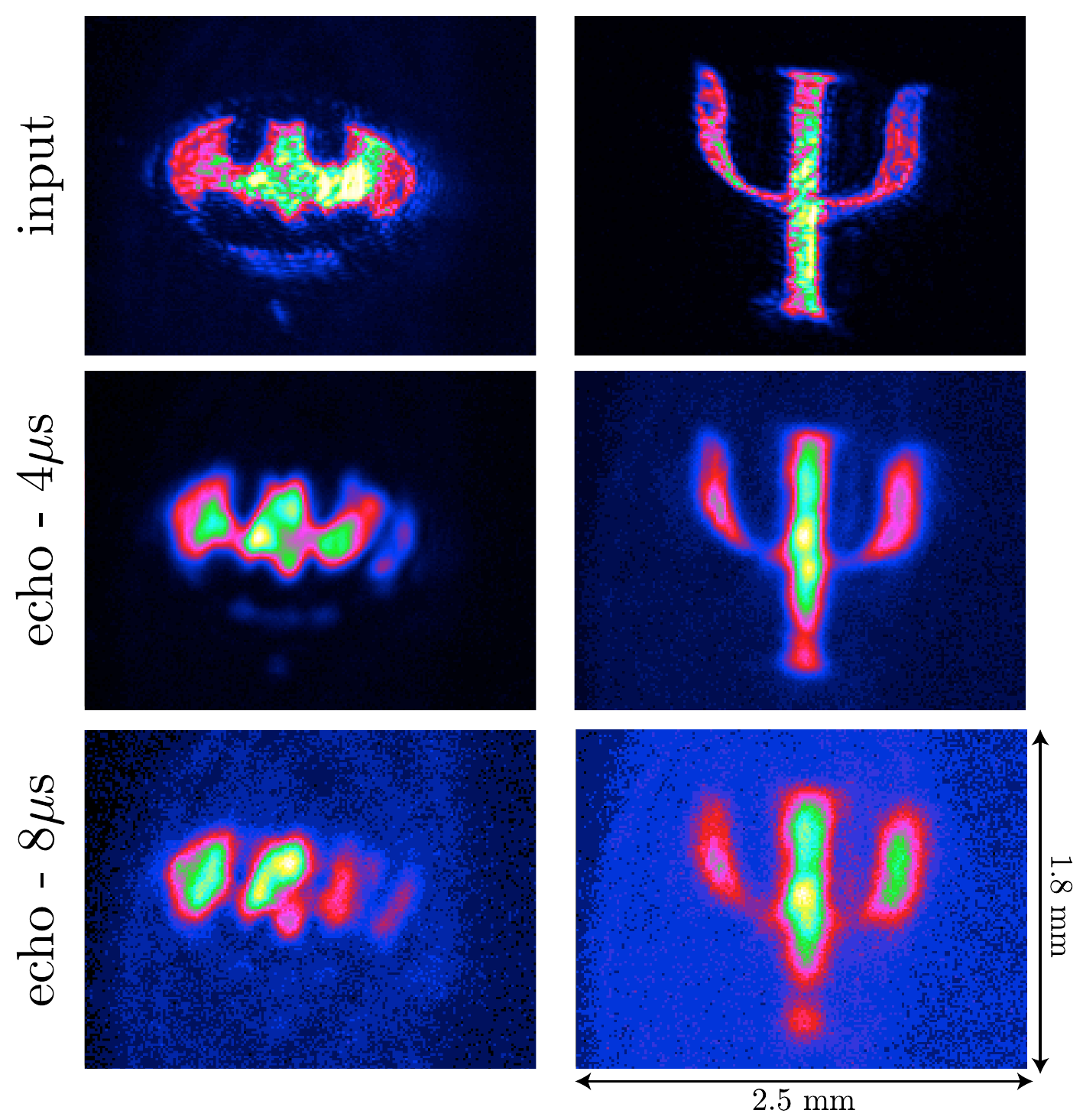}
\caption{Images stored in GEM: Bat signal (left), $\Psi$ (right). All images have been normalised to the same intensity scale. Recall efficiency $\approx 4\%$ at $4$ $\mu$s. }\label{images}
\end{minipage}
\end{figure*}

Our control station sampling rate limits trigger control precision to $1$ $\mu$s. This restricts the storage times for which we could take data and causes leading-edge clipping on some echoes. The effect of this clipping is to alternately reduce and increase the measured intensity. The effect is periodic, every three or four points the echo aligns with the optimal control window. The data is plotted with a three point smoothed function (dashed blue curve) which counteracts the clipping and an exponential fit to the dataset (blue curve). The red lines are the TEM-00 exponential fit decay curve normalised to the first measured efficiency of the higher order mode echoes, it is included to aid comparison between the figures.

We observe decay rates that vary with the phase complexity of the spatial mode, the more complex the mode the faster the efficiency loss. Additionally, the decay rate is correlated with the distance between out of phase peaks in the spatial mode of the input and output fields, consistent with the effects of spin wave diffusion under the model of equation \ref{recallsolution}. 

There are other possible explanations for the mode-decay correlation observed. Magnetic field variance off the beam axis may become significant as mode size increases and will further reduce recall efficiency. Longitudinal diffusion should also cause super-exponential decay in this storage scheme.

The Hermite-Gauss modes are not in general self similar under diffusion \cite{Firstenberg2010}, and the changing spatial intensity distribution provides additional evidence that spin wave diffusion is the dominant source of decoherence during storage. The TEM-20 mode, for example, has a small central lobe that is out of phase with the outer peaks.ÊWe expect, therefore, that the central lobe will be particularly prone to diffusion of atoms from surrounding regions.ÊIf we model the decay of the central lobe peak as a function of time (including both diffusion and inhomogeneous control field scattering during the read and write operations) we find that it decays faster than the outer peaks, as shown by the red line in Fig.~\ref{peak height}. ÊThis is a good fit to the experimental data, which although noisy, also appears to show accelerated decay of the central lobe. Transverse magnetic field inhomogeneity would, in contrast, lead to greater decay of the outer lobes.

We have observed that the Hermite-Gauss TEM modes do not decay identically. Furthermore, diffusion is a likely cause of the variation in the overall decay rate of the modes, and can be used to explain some of the finer structure of the decay curves with reference to the model introduced in Section IV-B.

\subsection{Spatially selective recall}

We demonstrated spatially selective recall and storage by aligning the control beam onto a single side of the TEM-10 spatial profile. The control beam makes possible the storage and re-emission of the illuminated half of the spatial profile. The dark-control component of the signal profile passes through the ensemble without interacting with the memory. When the atomic excitation rephases the illuminated side of the profile is released as a photon echo.

Figure~\ref{wink} shows the input and echo profiles for the case where we have selected the left and right halves of the profile. This process is a rudimentary example of a spatially selective memory operation. In this experiment only a single half of the profile was stored and recalled. Spatial multiplexing in which components of a stored profile are recalled separately using only in-memory operations is feasible with existing beam shaping techniques. Spatial mode operations of this sort are the basis of proposed multimode quantum networks \cite{Armstrong2012} that utilise optically entangled co-propagating spatial modes \cite{Janousek2009}.  

\subsection{Image storage}

The final result we shall present is a demonstration of multimode image storage in $\Lambda$-GEM. We present three images stored for up to $8$ $\mu$s and analyse the operation of the memory as a spatial frequency low pass filter with a steep frequency response.

In the previous sections the spatial profiles stored were pure Hermite-Gauss TEM modes generated by misalignment of a ring cavity. In order to store an image consisting of several co-propagating modes it is necessary to remove the redundant mode cleaner cavity and instead produce the desired image by passing the probe through a transparency mask which shapes the transmitted profile. 

In the previous configuration the probe and control beams were combined on the face of the mode-cleaning cavity exit port. For these results we combined the shaped probe beam with the Gaussian control beam on a non-polarising beam splitter. Naturally this sacrifices power in both beams, but we are able to compensate with power from the unnecessary cavity locking beam.

The control beam was the same Gaussian beam used in the previous experiments. It may be more efficient to optimise the control beam shape for the input image, or to shape the combined beam before the memory. However, for practical applications of image storage in multiplexed quantum repeaters it is important that the memory operate without prior knowledge of the input pulse.

The images and echoes are shown in Fig.~\ref{images}. We found the most successful image storage with a probe beam profile half the diameter of the beam used for single mode storage. At this beam size the images are close to the resolution limit of the CCD. The storage efficiency achieved is very low for both images, at $4$ $\mu$s the efficiency was approximately $4\%$. The spatial frequency components decay rapidly, consistent with the action of the diffusion Gaussian as a low pass filter. The diffusion timescale is substantially reduced by the smaller beam.

\section{Conclusions}

Our interest in multimode storage is driven by the desire for multiplexed quantum memories. These are essential for the construction of high bit-rate quantum repeaters for long distance quantum cryptography. Gradient echo memories with three-level atomic ensembles have the advantage of large spatial mode capacity and unique signal processing capability. Using a fast-triggering camera we have demonstrated the storage of multiple spatial modes in $\Lambda$-GEM, as well as the storage of multimode images, and investigated the factors limiting spatial fidelity. 

Brownian motion within the warm rubidium memory cell is the dominant cause of spatial decoherence during control-off storage. We describe a model of diffusion in the memory and show how diffusion causes the atomic spin wave to expand during storage. We measure this effect as broadening of the recalled photon echo. The mode expansion rate we measure corresponds to a diffusion coefficient of $D = 13.2$ cm$^2$/s, smaller than both theoretical predictions an independent experimental measurement.

Inhomogeneous control field scattering is the dominant cause of spatial decoherence during control-on storage. The two-photon scattering rate depends exponentially on the local control field power and therefore the control field profile contributes to loss of spatial fidelity over time. This effect is present during the read and write operations and needs to be considered during any spatial memory operations.

We demonstrated the storage of several higher order spatial modes with complex phase profiles and observed profile dependent efficiency loss rates. Diffusion between regions of the atomic excitation with opposite phase causes interference and additional efficiency loss consistent with our measurements. In any warm vapour memory diffusion of this sort poses a fundamental limit on the storage time for which a given profile is topologically stable.

Finally, we performed an in-principle demonstration of in-memory spatial operations using the control field. Spatially dependent storage and recall is merely the most basic of the spatial operations that are possible using $\Lambda$-GEM. The development of fast control beam shaping and transverse field control will provide additional degrees of freedom for spatial multiplexing and processing in GEM.

\begin{acknowledgments}
We would like to thank Quentin Glorieux for enlightening discussions during his visit to the ANU, and Joseph Hope and Brianna Hillman for sharing their preliminary results simulating diffusion in GEM. This research was conducted by the Australian Research Council Centre of Excellence for Quantum Computation and Communication Technology (Project number CE110001027).
\end{acknowledgments}

\end{document}